\documentclass{article}[12pt]
\usepackage{amsmath}
\usepackage{amssymb}
\usepackage{amsthm}
\usepackage{url}
\usepackage[figuresright]{rotating}

\usepackage{graphicx}

\input xy
\xyoption{all} \CompileMatrices

\DeclareMathOperator{\Prob}{Prob}

\newcommand{\A}{\mbox{\tt A}}
\newcommand{\C}{\mbox{\tt C}}
\newcommand{\G}{\mbox{\tt G}}
\newcommand{\T}{\mbox{\tt T}}
\newcommand{\gap}{\mbox{\tt -}}
\newcommand{\cX}{\mathcal{X}}
\newcommand{\cI}{\mathcal{I}}

\newtheorem*{DPP}{Dynamic Programming Principle}

\begin{document}

\title{Parametric inference of recombination\\in HIV genomes}

\author{Niko Beerenwinkel,
Colin N.\ Dewey,\\and Kevin
  M.\ Woods}


\maketitle

\begin{abstract}
Recombination is an important event in the evolution of HIV.  It
affects the global spread of the pandemic as well as evolutionary
escape from host immune response and from drug therapy within single
patients. Comprehensive computational methods are needed for
detecting recombinant sequences in large databases, and for
inferring the parental sequences.

We present a hidden Markov model to annotate a query sequence as a
recombinant of a given set of aligned sequences. Parametric
inference is used to determine all optimal annotations for all
parameters of the model. We show that the inferred annotations
recover most features of established hand-curated annotations. Thus,
parametric analysis of the hidden Markov model is feasible for HIV
full-length genomes, and it improves the detection and annotation of
recombinant forms.

All computational results, reference alignments, and C++ source code
are available at \url{http://bio.math.berkeley.edu/recombination/}.
\end{abstract}


\section{Introduction}
Retroviral recombination is a significant contributor to genetic
variation in HIV-1 genomes \cite{Robertson1995}.  When an individual
is infected by two or more different strains of HIV-1, recombination
can yield new forms of the virus that are mosaics of the original
strains. Given a particular viral genome, we would like to determine
whether it was formed by recombination and, if so, what
\emph{parental strains} recombined to form it.  This determination
is important in studying the geographic epidemiology of HIV-1
\cite{McCutchan2000}, and in the intra-patient evolution of immune
escape and of drug resistance in response to therapy
\cite{Kellam1995,Yusa1997}.  In this paper, we present a method for
inferring the parental strains of a recombinant genome.  This method
relies on a particular hidden Markov model (HMM) and involves
statistical inference for \emph{all} choices of model parameters.

Many models have been suggested for \emph{parental inference}, i.e.,
for the identification and characterization of recombinant
sequences. All rely, at some point, on a set of parameters. However,
small differences in the parameter values may result in
substantially different predictions by the model.  This is also true
of the model we present, but we will use the technique of
\emph{parametric inference} to ensure that we have a complete
understanding of how choices for the parameter values affect
predictions.

Given a multiple alignment of a recombinant viral genome with its
possible parental genomes, we use a probabilistic HMM to predict,
for each position in the recombinant genome, the parental strain
that gave rise to it. Our HMM is motivated by the \emph{copy-choice
model} \cite{Coffin1979}, which gives one possible biological
mechanism for viral recombination. This model is based on the fact
that recombination in retroviruses results from two RNA molecules
being packaged in one virion \cite{Hu1990}.  In multiply infected
cells, two distinct strains may be packed into a single virion
\cite{Jung2002}.  In the copy-choice model, a mosaic DNA molecule
results from reverse transcriptase jumping between two different RNA
templates during reverse transcription in the subsequently infected
cell.

We analyze our HMM over all choices of parameters using the
parametric inference technique, which is a general method for
evaluating graphical models. For example, parametric analysis has
been used successfully for pairwise sequence alignment
\cite{Gusfield1994,Pachter2004b,Waterman1992,Zimmer1997}. In this
paper, we describe the methods of parametric inference as they are
applied to our recombination HMM.  To evaluate our parametric
method, we use it to infer the parental subtypes of HIV-1
circulating recombinant forms (CRFs) obtained from the Los Alamos
HIV Sequence Database.

We show that a simple HMM, combined with the ability to evaluate the
model over its entire parameter space, is effective in predicting
parental subtypes for recombinant HIV-1 genomes.  We identify the
range of parameter values that maximize the accuracy of our
predictions on the test set, where we measure accuracy based on
concurrence with hand-curated annotations. We demonstrate that
parental subtype inference over all parameter values is feasible for
HIV full-length genomes, and that it is much more informative than
restricting to a particular choice of parameters.

\subsection{Related work}

There are many methods for the identification and characterization
of recombinant sequences (for a current list, see
\texttt{http://bioinf.man.ac.uk/robertson/\\
\~{}recombination/}). Most of these methods take as input a multiple
alignment of a putative recombinant with a collection of parental
sequences.  These methods either output a list of parental sequences
giving rise to each query position or simply state whether or not
the query is a recombinant.

One method that attempts to identify the parental subtype for each
position within a putative recombinant sequence was introduced in
\cite{Hein1993}.  This method models each column of the input
alignment by a hidden state that represents the tree topology of the
column.  Given fixed costs for substitutions and recombination
events (changes in tree topology), the algorithm gives a most
parsimonious explanation for the query sequence.  This model was
later formulated probabilistically as an HMM \cite{McGuire2000}.
Parameter estimation and extensions for this model have been studied
in depth \cite{Husmeier2001,Husmeier2005}.  These HMMs are similar
to multiple alignment methods that have incorporated the concept of
recombination \cite{Kececioglu1998,Lee2002}.  Indeed, our model can
be regarded as a specialization of ``jumping alignments'' that were
introduced for remote homology detection \cite{Spang2002}.

Several methods that have been applied to HIV recombinant forms use
a sliding window technique
\cite{Oliveira2005,MaynardSmith1992,Paraskevis2005,Siepel1995,Salminen1995}.
In this approach, subsequences of a certain fixed length are
considered along the genome and a test statistic is calculated from
each segment. These methods are local in the sense that they do not
solve a global optimization problem, but detect changes of locally
computed characteristics along the genome.

Methods like those just mentioned and others require many
parameters, which must be estimated.  Although many of these methods
can be effective, their performance is highly dependent on their
parameter values and the rates of evolutionary events that formed
the input sequences \cite{Posada2001b}.  In addition, the nucleotide
substitution models for these methods are usually fixed, even though
different models may be better for different data sets
\cite{Posada2001a}.  The framework of Bayesian statistics offers a
way to deal with uncertainty in model parameters, albeit at
considerable computational cost due to Markov Chain Monte Carlo
(MCMC) methods that are needed for estimating the posterior
probabilities~\cite{Suchard2002}.

We remedy parameter uncertainty by determining all solutions for
all parameter choices. For the presented model, a specific
parametric analysis was first conducted in \cite{Maydt2005} for the
detection of recombinant sequences.  Employing cost minimization,
the parametric solutions were analyzed in order to identify features
that indicate recombination.  Here, we take a probabilistic point of
view and present parametric inference methods that generalize
readily to more complex probabilistic graphical models.  Even though
the algorithms we present are quite general and return complete
information about dependence on parameter choices, they can master
full-length HIV genomes of about 10,000 base pairs.


\section{Methods}

\subsection{Hidden Markov model and dynamic programming}

We consider sequences over the 5-letter alphabet $\Sigma' = \{ \A,
\C, \G, \T, \gap \}$ of nucleotides supplemented by the gap
character. For a given multiple alignment $\mathcal{A}$ of $n$
columns and $N+1$ DNA sequences $y, s^{(1)}, s^{(2)}, \dots, s^{(N)}
\in \Sigma'^n$, the task is to explain the distinguished sequence
$y$ as a recombinant of the remaining sequences $s^{(i)}$, $i \in
\{1,\ldots,N\}$.  An \emph{annotation} of $y = (y_1,\ldots,y_n)$ is
a sequence $x=(x_1,\ldots,x_n)$ over the alphabet $\Sigma =
\{1,2,\ldots,N\}$, where $x_j = i$ signifies that the $j$th
character of $y$ originates from the $j$th character of $s^{(i)}$.
We introduce an HMM for inferring unobserved annotations from
observed sequences. An \emph{explanation} is an annotation that
maximizes the {\em a
  posteriori} probability of the data, given fixed model parameters.
Since the ``true'' values for the parameters are not known with any
certainty, we present a parametric analysis that yields all maximum
{\em a posteriori} (MAP) annotations for all parameter values.

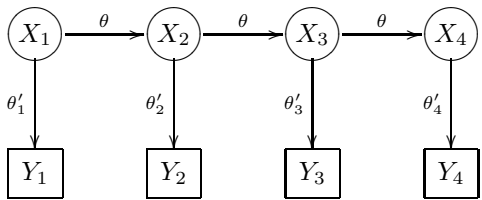
\begin{figure}[!tpb]
\centerline{
\begin{tabular}{c}
\xymatrix@!=1cm{
      *+<10pt>[o][F-]{X_1}\ar@{->}[d]_{\theta_1'}\ar@{->}[r]^\theta
   &  *+<10pt>[o][F-]{X_2}\ar@{->}[d]_{\theta_2'}\ar@{->}[r]^\theta
   &  *+<10pt>[o][F-]{X_3}\ar@{->}[d]_{\theta_3'}\ar@{->}[r]^\theta
   &  *+<10pt>[o][F-]{X_4}\ar@{->}[d]_{\theta_4'}  \\
      *+<10pt>[F]{Y_1}
   &  *+<10pt>[F]{Y_2}
   &  *+<10pt>[F]{Y_3}
   &  *+<10pt>[F]{Y_4}
}
\end{tabular}
} \caption{Graph of the recombination hidden Markov model for $n=4$
  alignment columns.  For column $j$ of a multiple alignment, the
  observed random variable, $Y_j$, is the character of the recombinant
  sequence for that column, and the hidden random variable, $X_j$,
  represents the parental sequence from which that character was
  derived.} \label{fig:HMM}
\end{figure}

The HMM has hidden random variables $X = (X_1,\dots,X_n)$, encoding
annotations in $\Sigma^n$, and observed random variables
$Y = (Y_1,\dots,Y_n)$, encoding sequences in $\Sigma'^n$.
The underlying graph of the model is depicted in Figure~\ref{fig:HMM}.

The dynamics of the model are given by the $N \times N$ transition
matrix $\theta$, which is the same for all transitions, and the $N
\times 5$ emission matrices $\theta'_j$, for $1\le j\le n$.  The joint
probability of an observed sequence $Y = y$ and an annotation $X = x$
is
\[
   f_{y,x} = \Prob(Y=y,X=x)
   = \pi_{x_1} \theta'_{1,x_1,y_1}
     \prod_{j=2}^{n} \theta_{x_{j-1},x_{j}} \theta'_{j,x_{j},y_j},
\]
where $\pi_i = \Prob(X_1 = i)$ is the initial distribution of $X_1$.

The likelihood of an observed sequence $Y = y$ is obtained by
marginalization,
\[
   f_{y} = \Prob(Y=y) = \sum_{x\in\Sigma^n} f_{y,x}.
\]
The sum can be evaluated efficiently due to the factorization
\begin{eqnarray*}
   f_y = \sum_{x_1\in\Sigma} \pi_{x_1} \, \theta'_{1,x_1,y_1}
     \Bigg( \sum_{x_2\in\Sigma} \theta_{x_{1},x_{2}} \, \theta'_{2,x_{2},y_2}
     \Bigg( \sum_{x_3 \in \Sigma} \cdots \\
   \qquad \cdots
     \Bigg( \sum_{x_n\in\Sigma} \theta_{x_{n-1},x_{n}} \, \theta'_{n,x_{n},y_n}
     \Bigg) \cdots \Bigg)\Bigg),
\end{eqnarray*}
which yields the forward algorithm. The factorization is valid
exactly because multiplication distributes over addition, a property
characteristic of \emph{semirings}. We have the following
generalization.  \pagebreak

\begin{DPP}
Let $\mathcal{S}$ be any set equipped with two binary operations
$\oplus$ and $\odot$ such that $(\mathcal{S}, \oplus, \odot)$ forms
a semiring. Let $\pi$, $\theta$, and $\theta'_j$ be matrices over
$\mathcal{S}$. Then the element
\begin{equation}   \label{eqn:f}
\bigoplus_{x\in\Sigma^n} \pi_{x_1} \odot \theta'_{1,x_1,y_1}\odot\bigodot_{j=2}^{n}\left(\theta_{x_{j-1},x_j}\odot\theta'_{j,x_j,y_j}\right)
\end{equation}
can be computed by the factorization
\begin{eqnarray}   \label{eqn:factorization}
  \bigoplus_{x_1\in\Sigma} \pi_{x_1} \odot \theta'_{1,x_1,y_1}\odot
    \Bigg( \bigoplus_{x_2\in\Sigma}\theta_{x_{1},x_{2}}\odot\theta'_{2,x_{2},y_2}
    \odot\Bigg(\bigoplus_{x_3\in\Sigma} \cdots \nonumber \\
  \cdots
    \Bigg(\bigoplus_{x_n\in\Sigma}\theta_{x_{n-1},x_{n}}\odot\theta'_{n,x_{n},y_n}
    \Bigg) \cdots \Bigg)\Bigg).
\end{eqnarray}
\end{DPP}
\noindent For variations and generalizations of this principle, see e.g.\
\cite{Aji2000,Bellman1957,Cowell1999,Durbin1998,Giegerich2004,Kschischang2001,Pachter2005}.
We will revisit the Dynamic Programming Principle several times,
with different semirings.

The explanations of $y$ are exactly the annotations $x$ satisfying
\begin{equation}   \label{eqn:MAP}
   f_{y,x} = \max_{x' \in \Sigma^n} f_{y,x'}.
\end{equation}
This condition on $x$ remains unchanged when passing to the logarithms
of the parameters $\pi_i$, $\theta_{i,i'}$, $\theta'_{i,j,k}$. Then the maximum
(\ref{eqn:MAP}) can be computed efficiently using the factorization
(\ref{eqn:factorization}) in the $(\max,+)$ algebra. This procedure is
exactly the classical Viterbi algorithm.
The $(\max,+)$ algebra
is also known as the tropical semiring, denoted
$(\mathbb{R} \cup \{-\infty\}, \max, +)$, and
solving (\ref{eqn:MAP}) is equivalent
to evaluating the polynomial (\ref{eqn:f}) in tropical arithmetic
\cite{Pachter2004a}.

\subsection{Model specification}

Motivated by the copy-choice model of recombination, we customize the
general HMM by specializing the matrices $\theta$ and $\theta'_j$.  We
assume that sequences change over time by two mechanisms: mutation and
recombination. Furthermore, these evolutionary events occur uniformly
over the sequence with unknown probabilities $\mu$ and $\rho$ for
mutation and recombination, respectively. For simplicity, we neglect
insertions and deletions in our model and encode alignment gaps with
the additional character ``\texttt{-}''.  Given the multiple alignment
$\mathcal{A}$ and parameters $\mu$ and $\rho$, we define
\begin{eqnarray*}
   \pi_i &=& \frac{1}{N},
   \\
   \theta_{i,i'} &=& \begin{cases}
     1-(N-1)\rho  & \mbox{if } i = i', \\
     \rho         & \mbox{else,}
   \end{cases}
   \\
   \theta'_{j,i,k} &=& \begin{cases}
     1-4\mu  & \mbox{if } S^{(i)}_j = k, \\
     \mu     & \mbox{else,}
   \end{cases}
\end{eqnarray*}
for $i,i' \in \Sigma$, $k \in \Sigma'$, and $j = 1,\dots,n$.

MAP estimation in this HMM is equivalent to optimizing the following
intuitive scoring scheme.  Given an annotation $x$, we identify two
salient features.  The first of these is $r$, the number of indices
$j$ ($1\le j\le n-1$) such that $x_j\ne x_{j+1}$, that is, the
number of recombination events. The second is $m$, the number of $j$
such that $y_j \ne s^{(x_j)}_j$, that is, the number of mutations
that must have occurred given that $y_j$ originated from the
sequence $s^{(x_j)}$.  We include in $m$ instances where one of
$y_j$ or $s^{(x_j)}_i$ is a gap character and the other is a
nucleotide.  Given choices for two real-valued parameters $R$ and
$M$, corresponding to recombination and mutation, respectively, the
\emph{annotation score} of $x$ is $R\cdot r+M\cdot m$.  We seek to
find the annotation $x$ that maximizes this score. The biological
meaningful solutions correspond to negative parameters $R$ and $M$.

We now show that this scoring scheme is equivalent to our hidden
Markov model. Given values for $R$ and $M$, let
\begin{equation}   \label{eqn:translation}
   \rho = \frac{e^R}{1+(N-1)e^R}\
     \quad \text{and} \quad
   \mu = \frac{e^M}{1+4e^M}.
\end{equation}
Note that $1-(N-1)\rho=\frac{1}{1+(N-1)e^R}$ and
$1-4\mu=\frac{1}{1+4e^M}$ are the other transition and emission
probabilities appearing in $\theta$ and $\theta'_j$.  Given an
annotation $x$ with $r$ recombination events and $m$ mutation
events, one can check that
\begin{eqnarray*}
 & &  \Prob(X=x,Y=y) \\
 & & \qquad = N^{-1} \rho^r\left(1-(N-1)\rho\right)^{n-1-r}\mu^m(1-4\mu)^{n-m} \\
 & & \qquad = \frac{e^{R\cdot r+M\cdot m}}{N \left(1+(N-1)e^R\right)^{n-1}\left(1+4e^M\right)^n}.
\end{eqnarray*}

Since the denominator is constant over all annotations $x$, and
since exponentiation is an increasing function, $\Prob(X=x,Y=y)$ is
maximized exactly when the score $R\cdot r+M\cdot m$ is maximized.

Conversely, given $\rho$ and $\mu$, let
$R=\log(\rho)-\log\left(1-(N-1)\rho\right)$ and
$M=\log(\mu)-\log(1-4\mu)$.  Since these are simply equations
(\ref{eqn:translation}) solved for $R$ and $M$, the most probable
annotation given by the HMM is exactly the annotation maximizing the
score $R\cdot r+M\cdot m$. Therefore our probabilistic HMM is
equivalent to the scoring scheme described above, and results can
easily be translated back and forth.  For the following parametric
analysis, the scoring scheme formulation is more natural.

\subsection{Parametric inference}

\begin{sidewaysfigure}
\includegraphics[width=\textwidth]{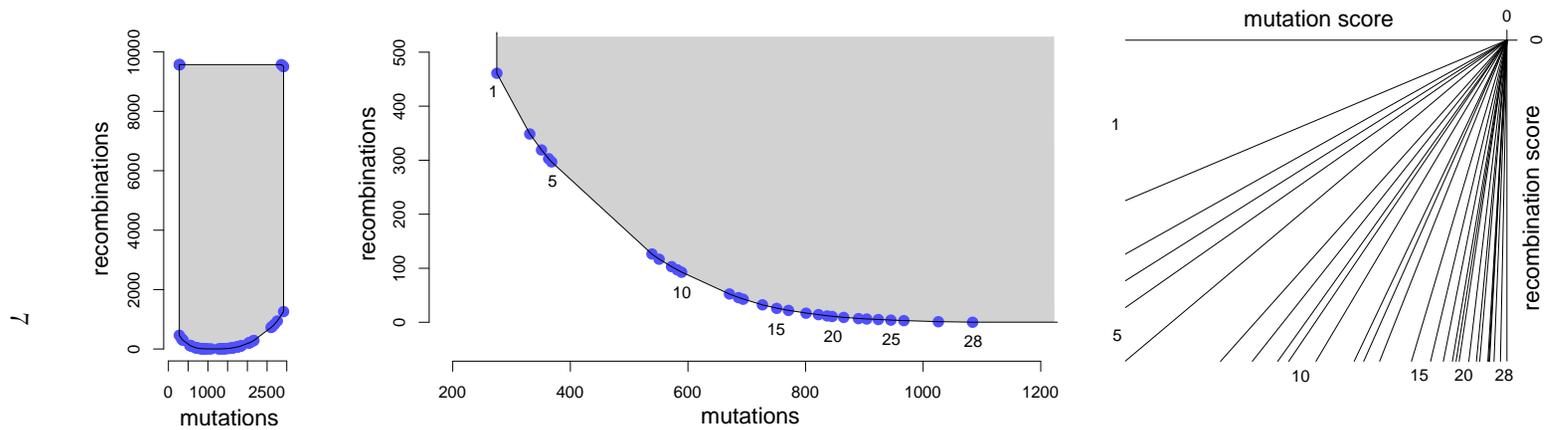}
\caption{Annotation polygon for the sequence
  {\sf CY.94.CY032\_AF049337}.
  Each point of the polygon represents all annotations with the same
  number of mutation and recombination events. The 72 vertices of the
  polygon correspond to explanations for different scoring schemes.
  Those 28 vertices corresponding to meaningful explanations are in the
  lower left corner of the polygon (on the left), which is shown in detail in
  the middle. The right panel displays the corresponding parameter regions for
  which each vertex is optimal (i.e., is an explanation).  }

\label{fig:polygon}
\end{sidewaysfigure}

Every annotation can be summarized by a pair $(r, m)$, where $r$ and
$m$ are the number of implied recombination and mutation events,
respectively.  Thus, all possible \emph{annotation summaries} can be
represented as a collection of points in the two-dimensional summary
space, with coordinates $r$ and $m$.  The {\em annotation
  polygon} is defined as the convex hull of this set of points
\cite[Sec.~3.2]{Pachter2005}.
Each point of the annotation polygon represents all annotations with
the same number of recombinations and mutations, and hence with the
same likelihood or annotation score.  Each vertex is the summary of
an explanation (optimal annotation), for some choice of parameter
values. The \emph{normal fan} of the annotation polygon is a
subdivision of the parameter space $(R,M)$ into a finite number of
regions (Figure~\ref{fig:polygon}). In this subdivision, the same
annotations are optimal for all choices of parameters that lie in
the same region.

The annotation polygon can be computed by the Dynamic Programming
Principle running in the {\em polytope algebra}. In our case, the
objects of this semiring are polygons, i.e., 2-dimensional
polytopes. Multiplication is defined as the Minkowski sum, and
addition as the convex hull of the set union. This algorithm is
known as {\em polytope propagation} \cite[Sec.~2.3]{Pachter2005}.

\subsection{Concurrence of explanations}

We want to compare inferred explanations $x$ with the ``true''
biological annotation $a$, which we take to be the hand-curated Los
Alamos annotation that is based on phylogenetic analyses
\cite{Robertson1997}.  This comparison can be done for all parameter
values of the model and hence allows for determining optimal
parameters.

Let $y$ be a sequence with true annotation $a \in \Sigma^n$.  When
comparing inferred annotations $x$ to the true annotation $a$, we
consider a rating scheme that encapsulates the most important aspect
of the annotation $x$, namely, whether it correctly states the set
of sequences that have recombined to form $y$. We define the
\emph{parental set} of $x$ as $I_x = \{i \mid x_j = i \text{ for
some }j\}$, i.e., the set of recombining sequences that $x$
indicates. The {\em concurrence} of a parental set $I$ to the true
annotation $a$ is defined as
\[
   c_a(I) = \frac{|I_a \cap I|}{|I_a \cup I|},
\]
and the concurrence of an annotation $x$ to the true annotation is
 $c_a(x)=c_a(I_x)$. Thus, if $x$ correctly names the
parental sequences, then $I_x = I_a$ and $c_a(x) = 1$. If $x$ and
$a$ have no recombining sequence in common, then $c_a(x) = 0$. In
general, a fixed choice of parameters will yield several optimal
parental sets. For a collection $\cX$ of annotations, we average the
concurrence $c_a$ over all sets $I$ that appear as a parental set
$I_x$, for $x \in \cX$. To be precise, define $\cI_{\cX}=\{I_x\mid
x\in\cX\}$, and let $c_a(\cX)$ be the average of $c_a(I)$ over all
$I\in\cI_{\cX}$. Note that we use an unweighted average: the number
of occurrences of a set $I$ as a parental set $I_x$ does not affect
it.

In order to rate all optimal annotations, we need to compute their
parental sets. This can be achieved by another instance of the
Dynamic Programming Principle. Let $\cI_{\cX}$ denote the collection
of parental sets that corresponds to a collection $\cX$ of
explanations. We consider ordered pairs $(\phi,\cI) \in \mathcal{S}
= \big(\mathbb{R} \cup \{-\infty\}\big) \times 2^{2^{\Sigma}}$,
consisting of  a number $\phi$ and a collection $\cI$ of subsets of
$\Sigma$.  For $\phi_1,\phi_2 \in \mathbb{R} \cup \{-\infty\}$ and
$\cI_1,\cI_2 \subset 2^{2^\Sigma}$, we define the operations
\[
   (\phi_1, \cI_1) \oplus (\phi_2, \cI_2) =
     \left( \max_{j=1,2} \phi_j,\:
     \bigcup\, \{ \cI_j \mid \phi_j=\max(\phi_1,\phi_2) \}
   \right)
\]
and
\[
   (\phi_1, \cI_1) \odot (\phi_2, \cI_2) =
     \left( \phi_1 + \phi_2, \:
     \{I_1 \cup I_2 \mid I_j \in \cI_j \} \right),
\]
which make $(\mathcal{S},\oplus,\odot)$ a semiring.

If we define the matrices $\pi$, $\theta$, and $\theta'$
over this semiring by setting
\begin{eqnarray*}
   \pi_i &=& \Big(0,\big\{\{i\}\big\}\Big)
   \\
   \theta_{i,i'} &=& \begin{cases}
   \Big(0,\big\{\{i'\}\big\}\Big) & \text{if $i=i'$,}\\
   \Big(R,\big\{\{i'\}\big\}\Big) & \text{else},\end{cases}
   \\
   \theta'_{j,i,k} &=& \begin{cases}
   \big(0,\{\emptyset\}\big) & \text{if $s^{(i)}_{j}=k$,}\\
   \big(M,\{\emptyset\}\big) & \text{else,}\end{cases}
\end{eqnarray*}
then the object defined by expression~(\ref{eqn:f}) is exactly the
pair $(\phi,\cI_\cX)$ with $\phi$ the optimal score and $\cI_\cX$
the corresponding collection of optimal parental sets. Thus, by
virtue of the factorization (\ref{eqn:factorization}), these
collections can be computed efficiently.

\begin{figure}[!tbp]
\centering
\includegraphics[width=5in]{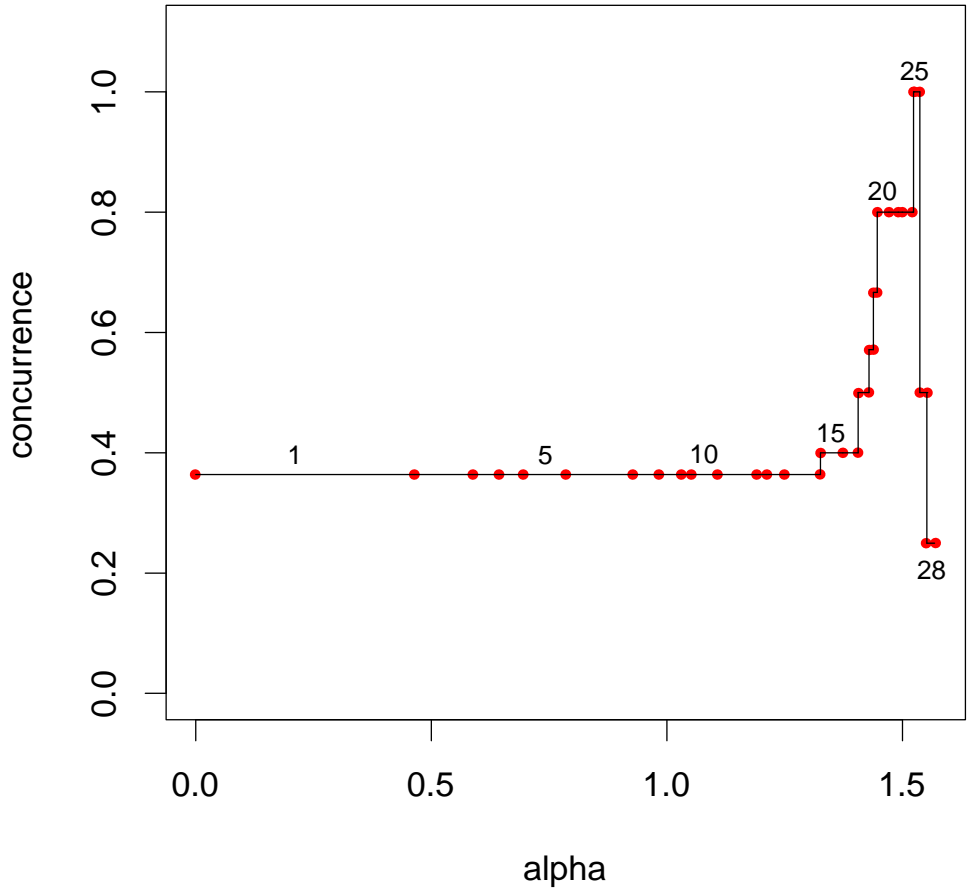}
\caption{Concurrence of model predictions with the true
  annotation for sequence {\sf CY.94.CY032\_AF049337}, over all parameter values.  For each value of the angle
  $\alpha = \tan^{-1} R/M$ ($x$-axis), ranges of which uniquely
  identify
  the biologically meaningful parameter regions in Figure~\ref{fig:polygon},
  the concurrence ($y$-axis) of the optimal
  parental sets with the true parental set is plotted.  MAP estimates
  remain constant over each parameter region (numbered
  according to Figure~\ref{fig:polygon}),
  which gives rise to a step function.}
\label{fig:score}
\end{figure}

\subsection{Data set}

We considered all of the HIV-1 full-length genomes in the Los Alamos
Sequence Database. A multiple DNA sequence alignment of length
12,635 was obtained from the 2003 HIV and SIV Alignments web site
(\url{http://hiv-web.lanl.gov/}).  We have omitted the CRF 01\_AE,
because of the lacking putative parental E strain. We further
excluded all recombinants that involve a CRF as one of the
recombining sequences. The resulting set of 341 genomes comprises 11
different subtypes and 11 different CRFs (Table 1).

Pure subtype sequences were used to build subtype consensus
sequences. The resulting consensus alignment was trimmed such that all
initial and terminal gaps were removed. The subalignment covers
positions 1,626 to 11,199 relative to the original alignment. These
positions correspond to 1,174 and 9,144, respectively, in the HXB2
numbering scheme \cite{Korber1998}.  By trimming, we effectively
excluded the flanking LTR regions, and small portions of the {\em gag}
(at the 3' end) and {\em nef} (at the 5' end) genes.

Thus, the data for each of the 341 recombination inference problems
consists of a multiple alignment of length 9,574 of one of the
sequences and the eleven consensus sequences.

\begin{table}[!t]
\label{tab:count} \centering
\begin{tabular}{lrclr}\hline Subtype & Count &
~~~~~~~~~~ & CRF & Count\\\hline
A1 & 38 &\phantom{---} & 02 (A,G) & 34\\
A2 & 4 &  & 03 (A,B) & 3 \\
B & 57 &  & 04 (A1,G,H,K) & 3\\
C & 109 & & 05 (D,F1) & 3\\
D & 39 & & 06 (A1,G,J,K) & 4\\
F1 & 6 &  & 07 (B,C) & 4\\
F2 & 4 &  & 08 (B,C) & 4\\
G & 7 &   & 10 (C,D) & 3\\
H & 3 &   & 12 (B,F1) & 5\\
J & 2 &   & 14 (B,G) & 6\\
K & 2 &   & 16 (A2,D) & 1\\\hline
\\
\end{tabular}
\caption{Analyzed HIV subtypes and circulating recombinant forms
(CRFs).}

\end{table}

\section{Results}

We have computed, for all 341 HIV-1 genomes, the annotation polygons
and all concurrence ratings for all parameter values of the HMM. The
Dynamic Programming Principle has been implemented in C++, and its
various occurrences have been realized using templates and operator
overloading, i.e., implementation of the respective semirings. The
annotation polygons can be computed in $O(N^2 n^{5/3})$ time
\cite{Pachter2004b}, and computing a typical HIV annotation polygon
takes less than one minute on a 2.6 Ghz Linux workstation. The
complete computational results and the source code are available at
\texttt{http://bio.math.berkeley.edu/\\recombination/}.

\subsection{Explanations}

An annotation polygon represents the set of all possible annotations
of a sequence as a recombinant of the 11 consensus sequences. For
example, Figure~\ref{fig:polygon} shows the annotation polygon for
sequence {\sf CY.94.CY032\_AF049337}, which is believed to be a CRF
04 comprising subtypes A1, G, H, and K in the Los Alamos database.

The annotation polygon induces a subdivision of the parameter space
into regions such that the same annotations are optimal for all
parameter values in a given region.  Figure~\ref{fig:polygon} shows
the biologically meaningful part of the polygon (middle) and the
corresponding subdivision of the parameter space (right) for all
non-positive $M$ and $R$ (i.e., where mutation and recombination is
penalized). Which annotations are optimal is uniquely determined by
the angle $\alpha = \tan^{-1}R/M$, and each parameter region is a
cone enclosed by a pair of rays. We refer to these parameter regions
by (intervals of) the angle $\alpha \in [0, \pi/2]$.

\begin{sidewaysfigure}
\centering
\includegraphics[width=\textwidth]{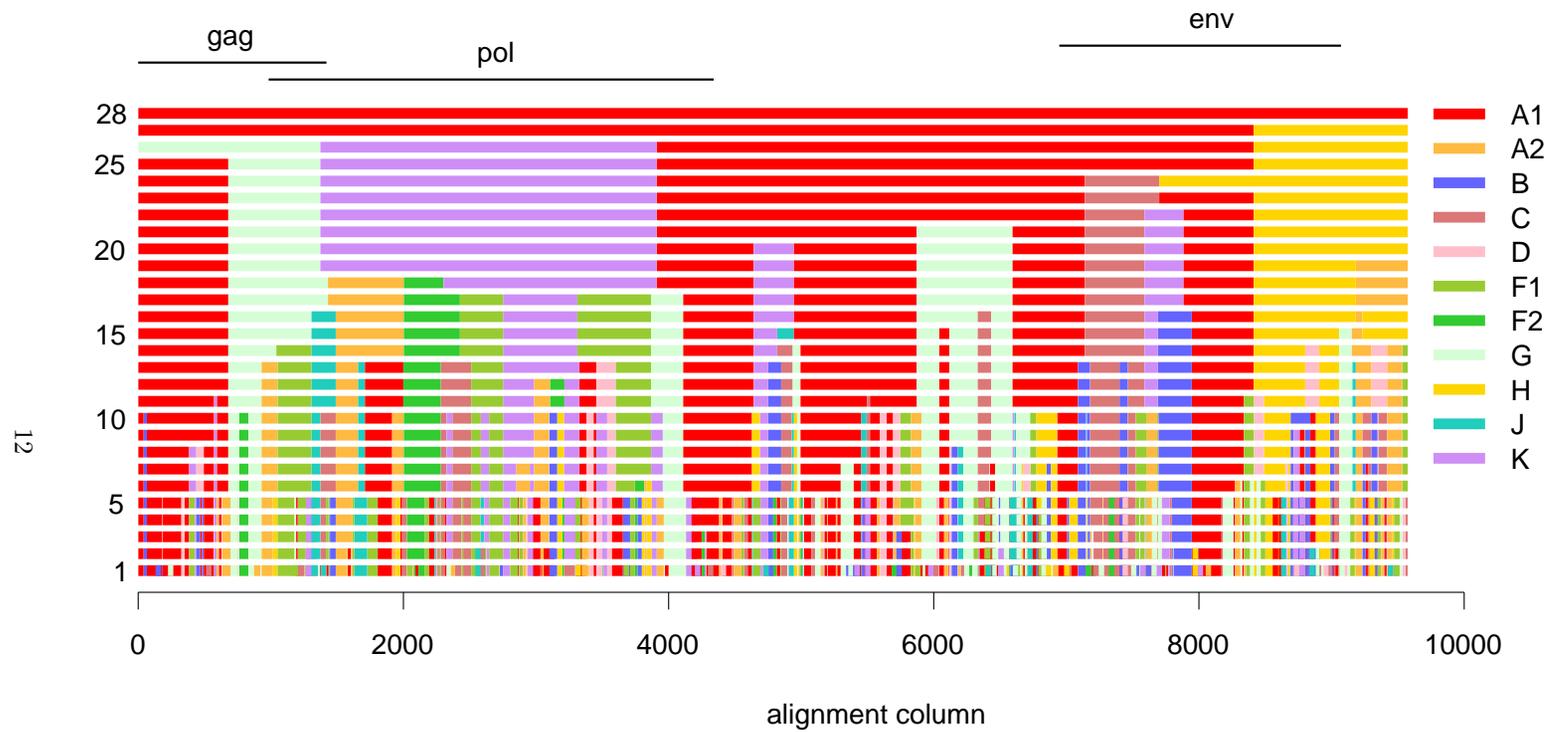}
\caption{Explanations (optimal annotations) for sequence {\sf
CY.94.CY032\_AF049337} (CRF 04, subtypes A, G, H, K), over all
meaningful parameter values. The numbers on the left correspond to
the labeled parameter regions in Figures~2 and 3. On top the
locations of the three major HIV genes are indicated. }
\label{fig:viterbi}
\end{sidewaysfigure}

For our analysis of the HIV-1 genomes, an annotation is a string
over $\Sigma = \{1,\dots,11\}$ of length $n = $ 9,574. In
Figure~\ref{fig:viterbi} we have picked one explanation for each
vertex in Figure~\ref{fig:polygon}.  From bottom to top, the
explanations correspond to parameter choices of increasing $\alpha$,
that is, of increasing ratio of the probability of a mutation to the
probability of a recombination. Explanations 25 and 26 use exactly
the set of recombining sequences A1, G, H, and K, although some
small sequence segments of the Los Alamos hand-curated annotation
are missing. However, most of these small segments appear in
explanations that correspond to smaller angles $\alpha$.  For
example, a subtype G fragment at the 3' end of the {\em pol} gene
appears in explanation 17 and below, and a subtype K segment in the
{\em env} gene occurs in explanations 22 to 17.  For explanations 24
and below, we estimate a segment in the 5' region of the {\em env}
gene to originate from subtype C, whereas in the Los Alamos
explanation this region is of unknown origin.  Thus, the parametric
solution of the inference problem recovers most features of the
established explanation and may serve as a starting point for new
sequences and for unexplained regions.

\subsection{Performance}

Computing the sets of recombining sequences for each vertex of the
polygon yields the concurrences for all parameter values in the
corresponding normal cones (Figure~\ref{fig:score}). Concurrences
are plotted against the angle $\alpha$. The resulting function takes
values in $[0,1]$, where 0 means that the parental sets of the
 predicted annotations and the true annotation have no sequence in
common, and 1 means that both sets coincide.  Since every parameter
value in a region yields the same set of explanations, the function
is a step function, constant within each region. For example, for
sequence {\sf CY.94.CY032\_AF049337}, the maximum score of 1 is
attained for the two subsequent parameter regions denoted 25 and 26,
which correspond to the two explanations labeled as such in
Figure~\ref{fig:viterbi}.

\begin{figure}[!tbp]
\centering
\includegraphics[width=4in,angle=270]{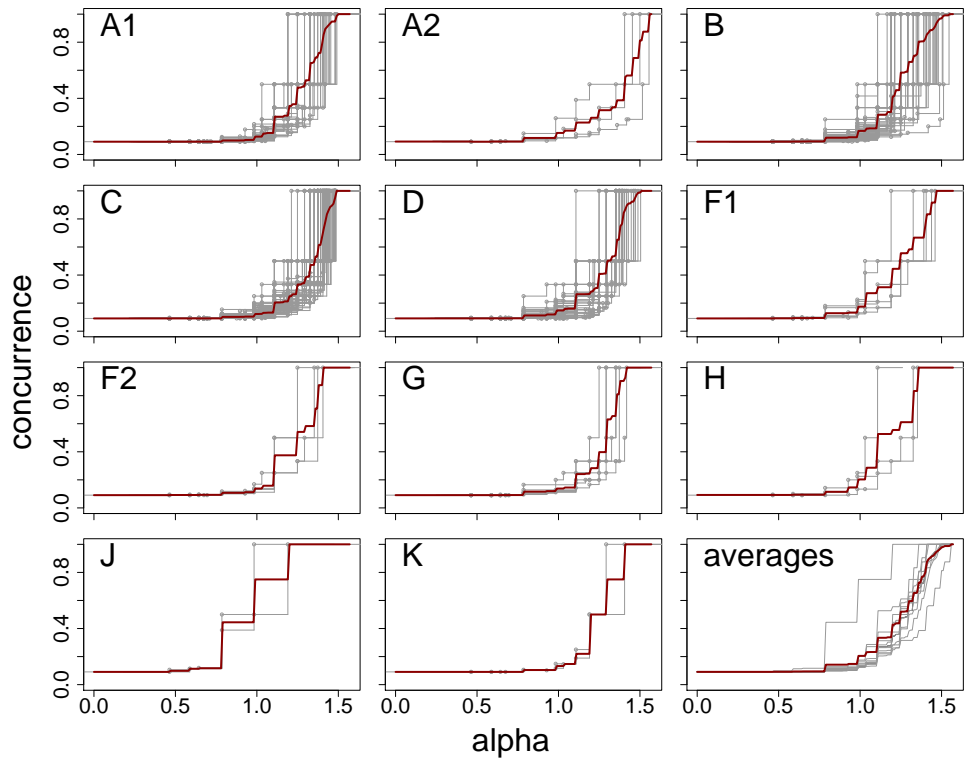}
\caption{Concurrences of model predictions with the true
  annotations for established pure subtype sequences, over all values
  of the model parameters.  Scores for individual sequences are
  plotted in light grey and average scores for all sequences of the same
  pure subtype is the thick, dark line.  The average curves for all of
  the subtypes are plotted together (grey) in the lower right
  plot, along with their average (dark).} \label{fig:pure}
\end{figure}

\begin{figure}[!tpb]
\centering
\includegraphics[width=4in,angle=270]{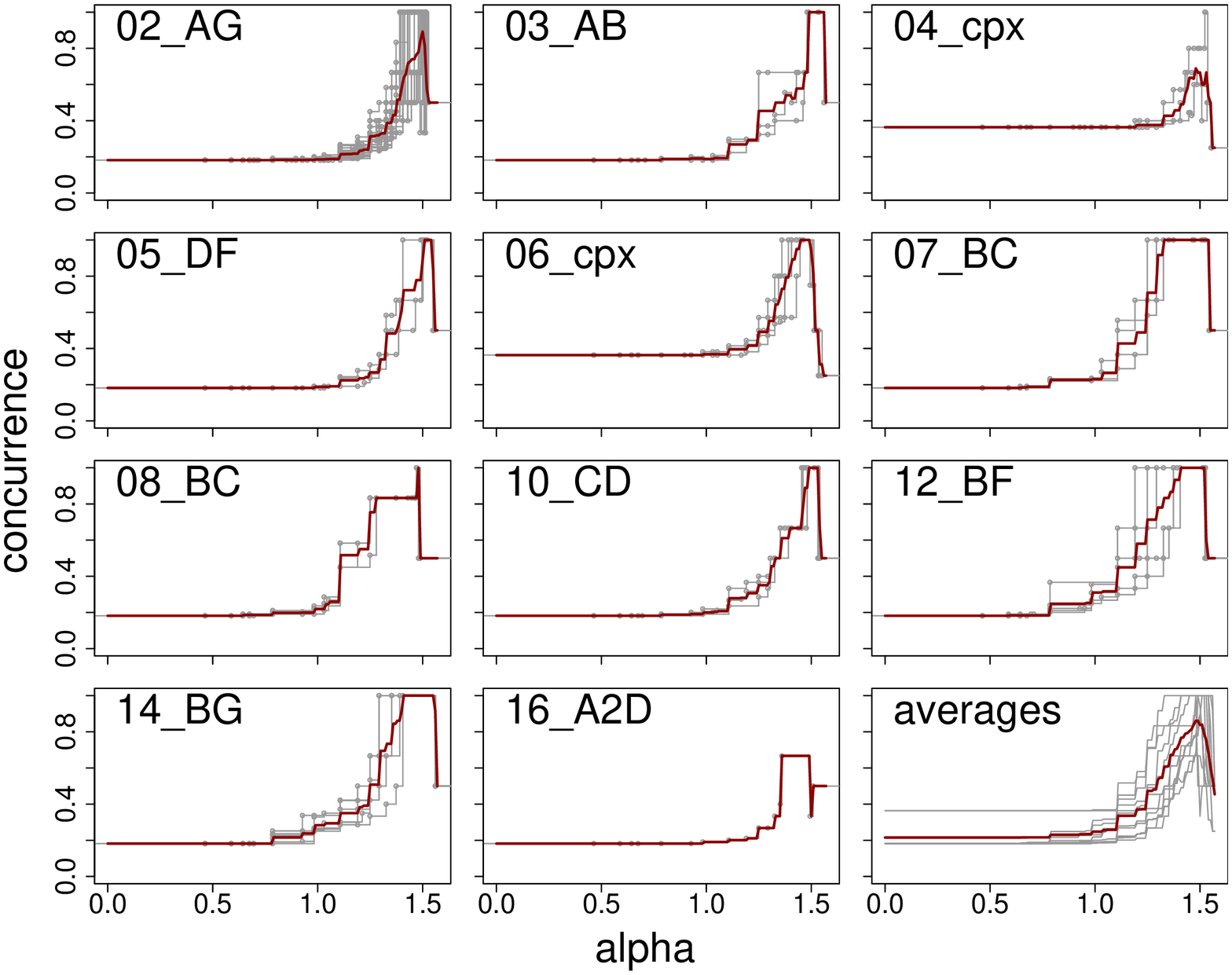}
\caption{Concurrences of model predictions with the true
  annotations for established CRF sequences.  Plots are analogous to
  those in Figure~\ref{fig:pure}.}
\label{fig:crf}
\end{figure}

We have computed the concurrences for all 341 genomes in order to
analyze the performance of the HMM in recovering the established
annotation from inferred explanations. In Figure~\ref{fig:pure} we
first analyze the 271 pure subtypes.

For $\alpha = \pi/2\approx 1.57$ all scores are 1, since this angle
corresponds to $\rho = 0$, i.e., recombination is impossible. In
fact, all $\alpha \ge 1.556$ yield the correct annotations for all
subtypes and so have a score of 1. The average scores per subtype
are displayed in the lower rightmost plot of Figure~\ref{fig:pure}.
For example, for $\alpha \ge 1.472$, the average of these averages
is still $\ge 0.95$. This average can be regarded as the expected
score when assuming uniform distribution of the subtypes.


For the CRFs, concurrence is no longer an increasing
function, because $\rho = 0$ cannot yield the true annotation
(Figure~\ref{fig:crf}). For many but not all CRFs, a score of 1,
i.e., perfect annotation, is reached.  The maximum expected score
under the assumption of uniform CRF distribution is 0.874 and it is
attained for $\alpha \in [1.481,\,1.485]$. This unique local and
global maximum is rather sharp indicating that the optimal value
does not vary a lot between different CRFs. Scores $\ge 0.85$ are
expected for $\alpha \in [1.472,\,1.494]$, a parameter region in
which the expected score for the pure subtypes is $\ge 0.95$.

\section{Discussion}

We have presented an HMM for the detection and annotation of
recombinant sequences. The unifying Dynamic Programming Principle
was applied for MAP estimation, parametric
inference, and validation via concurrence rating.
As a probabilistic model, the parameters $\rho$ and $\mu$ of the HMM
can be interpreted as the probability of observing a recombination
event and a mutation event, respectively, in the data set.
The HMM also allows one to compute the posterior
probability of each putative parent at each site of the genome.

The model can be extended in many directions, while maintaining the
same conceptual and computational framework.  The emission matrices
$\theta'_j$ may be parametrized to implement more flexible
substitution models than the Jukes-Cantor model (which we implicitly
assume), e.g., by accounting for transition/transversion bias. The
model might be expanded by explicit modeling of insertions and
deletions, thereby simultaneously searching for recombination and
aligning the query sequence to the reference alignment. This problem
can also be cast in an HMM.  The corresponding higher dimensional
polytopes and concurrence ratings for any of these models can be
computed efficiently using the Dynamic Programming Principle. In
addition, other more refined rating systems could be used to
determine the accuracy of the explanations returned by the model. On
the other hand, the power of our model to predict recombination, as
measured by concurrence ratings, is already quite high, and we need
only two parameters.

Parametric inference is very efficient for the presented model. We
have used it to analyze 341 HIV genomes of size approximately 10,000
characters. Furthermore, the 2-parameter model has a concise
representation of its set of explanations
(Figure~\ref{fig:viterbi}), and hence the parametric inference
solution can easily be inspected. Indeed, the case of HIV-1 genomes
has shown that the set of all solutions for all parameters can be
much more informative than a single estimate. For example, as
discussed in Section~3.1, certain parameter values may be
appropriate for determining the parental set of a sequence, whereas
other values may pick up shorter parental subsequences. All of this
information is inherent in the annotation polygon.

Solving the parametric inference problem should be regarded as an
offline precomputing step. Once its solution is available,
statistical inference for fixed parameters becomes very cheap. In
fact, MAP estimation in the HMM corresponds to solving a linear
program on the annotation polygon. In the absence of labeled
training data, i.e., annotated recombinants, the model parameters
may be estimated by maximum likelihood using the Expectation
Maximization algorithm. In those situations the annotation polygon
can be used to assess the sensitivity of MAP estimation with respect
to the uncertainty inherent in parameter estimation. The relative
position of the parameter estimate to the boundaries of the discrete
parameter regions determines the robustness of explanations.

Parametric analysis also provided the basis for evaluating the HMM.
We used the parental sets as the feature of an
explanation to be compared to the true annotation. Unlike the
collection of explanations for a given choice of parameters, the
collection of parental sets can be efficiently computed. For
example, there are 28,704 explanations for sequence {\sf
CY.94.CY032\_AF049337} in parameter region 25
(Figures~\ref{fig:polygon},\ref{fig:score},\ref{fig:viterbi}), but
all share the same single parental set \{A1, G,
H, K\}. Thus, restricting to parental sets and ignoring
recombination breakpoints is a compromise between computational
feasibility and accuracy.
The analysis of these concurrence
ratings has identified optimal parameter regions.
In practice, prior
knowledge about the mosaic structure of a sequence will affect the
choice of the parameter region to investigate.

In summary, annotation polygons provide a powerful tool
for inferring the recombinant structure of HIV genomes.

\section*{Acknowledgments}

Niko Beerenwinkel was supported by the Deutsche
Forschungsgemeinschaft under grant BE~3217/1-1, Colin Dewey by the
NIH (HG003150), and Kevin Woods by the NSF (DMS-040214).



\ \\

\noindent Niko Beerenwinkel, \textsc{Department of Mathematics,
University of California, Berkeley 94720, USA},
\texttt{niko@math.berkeley.edu}.

\ \\

\noindent Colin Dewey, \textsc{Department of Electrical Engineering,
University of California, Berkeley 94720, USA},
\texttt{cdewey@eecs.berkeley.edu}.

\ \\

\noindent Kevin Woods, \textsc{Department of Mathematics, University
of California, Berkeley 94720, USA},
\texttt{kwoods@math.berkeley.edu}.

\end{document}